\documentclass[12pt]{iopart}
\begin{document}

\title[]{ Wave particle duality: Merging de Broglie's ''double solution'' waves into (3+0)D electromagnetic solitons.}
\author{J. Moret-Bailly 
}
\address{Laboratoire de physique, Université de Bourgogne, BP 47870, F-21078 Dijon cedex, France.
email : jmb@jupiter.u-bourgogne.fr
}

\begin{abstract}
	The well known light filaments, are obtained in various media whose index of refraction increases before a 
saturation with the electric field; adding a small perturbation which increases the index  with the magnetic field, and 
neglecting the absorption, a filament curves and closes into a torus. This transformation of a (2+1)D soliton into a 
(3+0)D soliton shows the existence of those solitons, while a complete study, with a larger magnetic effect, would 
require numerical computations, the starting point being, possibly, the perturbed, curved filament.
	The flux of energy in the regular filaments is nearly a ''critical flux'', depending slightly on the external fields, so 
that the energy of the (3+0)D soliton is quantified, but may be slightly changed by external interactions.
	The creation of electron positron pairs in the vacuum by purely electromagnetic fields shows a nonlinearity of 
vacuum at high energies; supposing this nonlinearity convenient, elementary particles may be (3+0)D solitons.
	Set that the nearly linear part of the soliton is the $\Psi$ wave while the remainder, (the torus) is the $u$ wave; if 
the torus is not absorbed by Young slits, it is guided by the transmitted, interfering remainder of the $\Psi$ wave.

\end{abstract}
\maketitle
%

\section{Introduction}
As a linear combination of the solutions of a linear wave equation is, by definition, a solution, it is not possible to bind 
to a solution, without ambiguity or imprecision, a remarkable region called ''particle''; it is a source of paradoxes in 
quantum mechanics. On the contrary, in one dimensional, nonlinear problems, wave packets called (1+1)D solitons are 
stable, even when colliding with other solitons, and most of their energy remains in a limited, but propagating region of 
the space, so that they can be considered as particles. However, a true particle must be defined in three dimensions, and 
can be stopped: it must correspond to a (3+0)D soliton, but it seems that no such soliton was found up to now. The aim 
of the paper is transforming solutions of Maxwell's equations in nonlinear dielectrics, observed as light filaments, into 
(3+0)D solitons, and using the observed properties of these filaments which often cannot been demonstrated 
theoretically, to find properties of these solitons which allow to consider them as particles.

\medskip

The filaments of light which are (2+1)D solitons have been extensively studied \cite{ 
Chiao,Marburger,Stegeman,Kivshar} although the theory is difficult \cite{Feit} because the mathematics of nonlinear 
systems is poor. The required nonlinearity may be provided by many effects, in particular Kerr or photorefractive; here, 
we will consider the propagation of a monochromatic wave in homogenous, isotropic, media in 
which a nonlinear variation of the index of refraction is a function of the modulus of the amplitude of the electric field; 
this function may be similar to the function corresponding to a saturated Kerr effect, but it is only needed that it allows 
the filament solutions of the propagation equation to be very stable. With the assumed isotropy the electric field will be 
considered as a scalar or a vector.

In section 2 the interaction with external fields and the stability of the filaments are described, using observed properties 
rather than the theory which works only for some perfect geometrical configurations.

In the two next sections, the existence of an isotropic, non-absorbing, conveniently nonlinear medium is assumed. A 
(2+1)D filament soliton is transformed into a (3+0)D soliton by the introduction of a magnetic nonlinearity: after the 
setting of notations, it is shown than a small increase of the index of refraction by an increase of the modulus of the 
magnetic field may introduce an instability of the straightness and produce a curvature depending only on the 
properties of the medium; this curvature propagates, so that the axis of the filament becomes a circle.

Section 5 is a preliminary discussion of the interaction of such a soliton with Young holes or slits, assuming that 
vacuum has the convenient nonlinearities, and supposing that the curvature of the filament does not change much its 
interaction with the external fields. The nearly linear part of the field appears as a ''pilot wave'', de Broglie's $\Psi$ 
function while the remainder may be the $u$ wave.

\section{Stability of the filaments of light and interaction with fields.}
Assuming the cylindrical symmetry and a saturation of the nonlinearity, the flux of energy in a filament has ''critical 
values'' corresponding to modes, but we consider only the lowest energy solution, the other being unstable. Remark, 
although it is not optically realistic, that other stable solutions are found if the index of refraction has maximums while it 
increases, in the average, with the electric field.

The theory of nonlinear propagation is very difficult, so that many results are only found from the interpretation of the 
experiments.

The flux of energy in the filament may be absorbed strongly while the filament is long \cite{Brodeur}; to maintain its 
energy not far under the critical energy, to recover the lost energy, the filament absorbs the surrounding field: it is an 
equilibrium between the flux of energy in the filament, near the critical value, and the surrounding field; this absorption 
of energy by the filament is shown too by the birth of the filament in a wide laser beam. Conversely, if a single filament is 
produced by an homogenous enough laser beam with more than the critical energy, its diameter oscillates and it loses 
energy.

The large powerful laser beam which splits into filaments \cite{Brodeur} is far from being homogenous, but the filaments 
never merge, the filaments repulse each other; as the filaments recover energy from the surrounding field, the field is 
lower between two filaments than elsewhere: the repulsion shows that the filaments are attracted to the higher 
surrounding field.

After perturbations which have the cylindrical symmetry and do not change the energy, a stable filament recovers its 
initial properties except for a global phase change of the field. The observed stability implies that the filament is not 
destroyed by any small enough perturbation. In an homogenous, isotropic medium, the cylindrical symmetry is 
recovered; on the contrary, the experiments done in photorefractive media show that a lack of homogeneity of the 
excited medium curves the filament, while its sections perpendicular to the direction of propagation become nearly 
elliptical \cite{Petter}; the complicated interaction of a filament with an other filament \cite{Shih} curves the axis of the 
filament without destruction too.

\section{Mathematical settings}
Set $\mbox{\boldmath $E_0$}(x, y, z, t)$ the electric field, assumed polarised along $Ox$, of a stable 
filament propagating along the $Oz$ axis, in an homogenous, isotropic, lossless medium having a 
convenient isotropic electrical nonlinearity, and $E_0(x, y, z, t)$ the non-zero component of this field. 

\medskip

The electric field $\mbox{\boldmath $E_0$}(x, y, z, t)$ is an exact solution of Maxwell's equations in 
which the relative permittivity $\epsilon$ is a function of $|\mbox{\boldmath $E$}(x, y, z, t)|$, and the 
relative permeability $\mu$ is 1. Neglecting $|\mbox{\boldmath $\nabla . E$}(x, y, z, t)|$, Helmholtz 
propagation equation is obtained
\begin{equation}
\mbox{\boldmath $\Delta E$}(x, y, z, t)=\frac{\mu\epsilon(|E(x, y, z, t)|)}{c^2}{\partial^2 {\bf  E}(x, y, 
z, t)\over\partial t^2}.\label{propa}
\end{equation}
Set \cite{Chiao}
\begin{equation}
\mbox{\boldmath $E$}_0(x, y, z, t)= \mbox{\boldmath $E$}_t(x,y)\cos(\phi-\omega t) \label{propag}
\end{equation}
with $\phi=kz $.
$ \mbox{\boldmath $E$}_t(x,y)$ verifies the radial equation of the filament and may be written $ 
\mbox{\boldmath $E$}_t(r)$ while $\cos(kz-\omega t)$ is the propagation term. Assuming that the 
beam is cylindrical, and that there is no absorption, the time-reversal invariance shows that $ 
\mbox{\boldmath $E$}_t(x,y)$ is real, so that the wavefronts are planes perpendicular to $Oz$.

The energy propagates on light rays parallel to $Oz$, but the index of refraction depends on the field, 
thus on the radius $r$ and may be written $n(r)$; the optical path between two wavefronts is a 
decreasing function of $r$.

Remark that equation \ref{propag} which is a consequence of the assumed symmetries does not depend 
on the approximate computations.
\section{Perturbation by a magnetic nonlinearity.}
\medskip
Consider a slice of the filament between the planes $\Pi_0$ of equation $z=0$ and $\Pi_{\delta z}$ of 
equation $z=\delta z $.
Set $\Pi_\alpha$ the plane of equation  $z=\delta z (1+x\alpha)$, where $\alpha$ is a small constant; 
suppose that a daemon changes the index of refraction in the sheet, so that the phase of the field $\phi$ 
becomes a function of $x$ and $y$ getting in the plane $\Pi_\alpha$ the values it had without the 
perturbation, for the same $x$ and $y$, in $\Pi_{\delta z} $.

The component of the curl of the electric field along $Oy$ becomes $\delta E_0(x, y, z, t)/( \delta z(1+x\alpha))\approx(1-
x\alpha) \delta E_0(x, y, z, t)/ \delta z$, neglecting high order terms. Assuming that the index of refraction $n(x,y)$ 
increases up to a saturation with the modulus of $|\mbox{\boldmath $\nabla \wedge E$}_0(x, y, z, t)|$ , that is with the 
time derivative of the modulus of the magnetic field (or with the modulus of the amplitude of magnetic field for a fixed 
frequency), $n(x,y)$ becomes a growing function of $\alpha$ for $x$ positive.

The index of refraction is mostly changed by the perturbation where the electromagnetic field is large, so that one may 
think that the wave surfaces are not anymore plane; but the variations of index are very similar to the one which occur 
bending a filament in a photorefractive material: assuming the stability of the filament, the wave surfaces are not 
distorted by the perturbations, but odd terms in $\alpha$ must be considered in a development of $n(x,y)$, so that the 
wave surface for $z=\delta z $ is turned by an angle $\beta(\alpha)$. 

Adjusting the parameters of magnetic nonlinearity of the medium, $\beta(\alpha)$ may be equal to $\alpha$ for a certain 
$\alpha_0$, with a sign of 
the derivative which provides a stability: thus the problem is self-consistent without daemon. The value $\delta 
z/\alpha_0$ of the radius of curvature depends only on the properties of the medium. A small, local perturbation 
destroying the straightness of the filament, the perturbation propagates, the filament curves into a torus; adjusting the 
frequency, the output field after a turn is exactly the assumed input, the filament closes into a (3+0)D soliton. 

The flux of energy in the filament being near the critical value, and the length of filament transformed into a torus 
depending on the assumed properties of the medium, the energy of the soliton is quantified.

\medskip

The existence of  electromagnetic (3+0)D solitons has been shown, apparently for the first time, in the particular case 
where the variation of the index of refraction produced by the magnetic field is low; it may be a starting point for 
numerical computations of more general solitons; numerical computations seem necessary to answer many questions 
such as:

- what happens increasing the magnetic nonlinear contribution to the index of refraction: in particular, can the torus 
become next to a sphere ?

- can the soliton have an electric or magnetic charge introduced by a non zero divergence of the fields ?

The process of building (3+0)D solitons may be extended, for instance introducing a torsion.
\section{Solving the wave particle duality by solitons}
In the vacuum, up to X rays, Maxwell's equations with linear parameters are well verified, but a powerful enough 
$\gamma$ photon interacts with an electric or magnetic field, or with an other $\gamma$ photon to produce an electron 
pair \cite{Ritus}. In quantum physics, the required nonlinearity is introduced through virtual particles; in a classical 
scheme, nonlinear terms must be introduced in Maxwell's equations (Schwinger \cite{Schwinger}); this introduction 
breaks the superposition property of linear systems, gives individualities to regions of field such as (3+0)D solitons. The 
properties of the ring model of the electron \cite{Allen} bound to its symmetries could apply to the (3+0)D soliton.

\medskip

Split the electromagnetic field, somewhat arbitrarily, into two parts: the nearly linear part and the remainder identified 
respectively  to the $\psi$ and $u$ functions introduced by de Broglie \cite{deBroglie}. Show that the $\Psi$ wave is an 
''onde pilote'' :

Remark first that even if a good test of nonlinearity allows to split the soliton wave well, the $\Psi$ wave remains badly 
known: our solutions of Maxwell's equations are generally mathematically exact, physically wrong because, for instance, 
the absorption of a wave emitted by one dipole requires an infinity of dipoles: it remains everywhere the zero point 
electromagnetic field, the addition of which transforms the first Planck's law into the second. This field perturbs 
permanently the electrons (producing the Lambshift, for instance) and the soliton. The zero point field comes from a lot 
of sources, so that it is generally incoherent.

Suppose that the $u$ soliton, exactly as a filament, goes where the field is high. Hitting a screen pierced of two Young 
holes, the $u$ part of a soliton may be absorbed; if it is transmitted, the soliton has lost a part of its $\Psi$ wave and the 
remainder interfere; the soliton is not much perturbed by the incoherent part of the field so that it goes to a region of 
high field, that is a bright interference fringe: the $\Psi$ wave is a pilot wave.
\section{Conclusion}
Quantum electrodynamics is not necessary to compute the results of all experiments of optics (including fourth order 
interferences), provided that one uses the right classical electrodynamics which includes the zero point field. Assuming 
an electric and magnetic nonlinearity of vacuum at very high frequencies, it seems that this result extends to a classical 
interpretation of the interferences of massive particles considered as solitons of the electromagnetic waves.

\end{document}